\documentclass[slac_one]{revtex4}
\usepackage{graphicx}
\usepackage{fancyhdr}
\begin{document}

\title{High Mass Dijet and $\bf{t\bar{t}}$ Resonance Searches} 

%

\author{Alex Melnitchouk for the CDF and D0 Collaborations}
\affiliation{University of Mississippi, University, Mississippi, 38677, USA}
%

\begin{abstract}
We present searches for dijet and $t\bar{t}$ mass resonances using 
between 0.68 and 2.1 fb$^{-1}$ of Tevatron Run
II data collected by the CDF and D0 detectors. 
No evidence of new physics is found,  
and 95\% C.L. limits are set on a number of new physics hypotheses,
such as excited quark, Randal-Sundrum graviton, Z', W', color-octet
technirho, axigluon and flavor-universal coloron, E$_{6}$ diquark,
quark compositeness, ADD and TeV$^{-1}$-sized LED, massive gluon.
\end{abstract}

\maketitle

\thispagestyle{fancy}


\section{INTRODUCTION AND OVERVIEW OF ANALYSES} 
Of all high $p_T$ processes at a hadron collider, QCD processes
have the largest cross-section. Large data samples accumulated
by the CDF and D0 detectors in RunII allow precise measurements
of the shapes of the observables that describe Standard Model QCD processes.
Therefore deviations from the Standard Model could be easily detected.

The top quark, due to its large mass, that is not yet understood,
may have a special connection with the electroweak symmetry breaking
and new physics. At Tevatron the $t\bar{t}$ pairs are predominantly produced
via the s-channel gluon annihilation.
However this may not be the only production mechanism. 
A new neutral heavy particle produced in a proton-antiproton collision
would decay into $t\bar{t}$, adding a resonant component to the Standard Model spectrum.
An example of such particle is a heavy resonance, Z', which appears in theories
with a new strong gauge force coupling to third generation.
Besides, a top and anti-top quarks may form a bound state before decaying.
This would also create a bump in the $t\bar{t}$ invariant mass spectrum.

Six searches are presented in this paper, two analyses in the dijet final state
and four analyses in the $t\bar{t}$ final state. In the dijet final state,
CDF looked for a bump in the invariant mass spectrum with 1.13 fb$^{-1}$,
while D0 exploited the shapes of dijet angular distributions with 0.7 fb$^{-1}$.
In the $t\bar{t}$ final state there is one analyses by D0 with 2.1 fb$^{-1}$,
and three analyses by CDF with 0.96~\cite{cdf-prd}, 0.68~\cite{cdf-prl}, and 1.9 fb$^{-1}$ respectively.
Important aspect of the $t\bar{t}$ analyses is the reconstruction of top-antitop
invariant mass. D0 uses direct mass reconstruction, without constraint fit.
CDF analyses employ tempate method, matrix method, and dynamic likelihood method,
respectively. Analyses by both CDF and D0 $t\bar{t}$ use lepton+jets final state.

\section{DIJET SEARCHES}

\subsection{CDF Dijet Analysis}
\begin{figure*}[t]
\centering
\includegraphics[width=90mm]{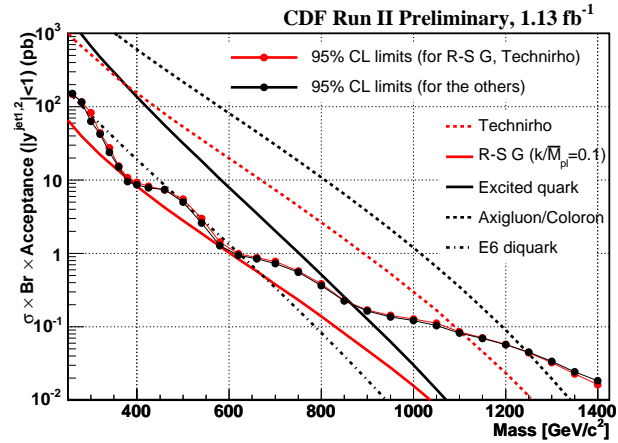}
\caption{Observed limits on resonant dijet production (CDF).} \label{cdf_dijet__3}
\end{figure*}

Jets are reconstructed with the midpoint cone 0.7 algorithm
within rapidity of 1.0. Events with dijet masses above 180 GeV
are selected. Cosmic event background is removed using missing
transverse energy significance cut. Dijet mass spectrum is fitted
with the parameterized shape that was derived from the full detector
simulation spectrum. No significant indication of resonant structure
is observed. Uncorrected data are used to set limits using Bayeasean
approach. The results are shown in Figure~\ref{cdf_dijet__3}.

\subsection{D0 Dijet Analysis}
\begin{figure*}[t]
\centering
\includegraphics[width=90mm]{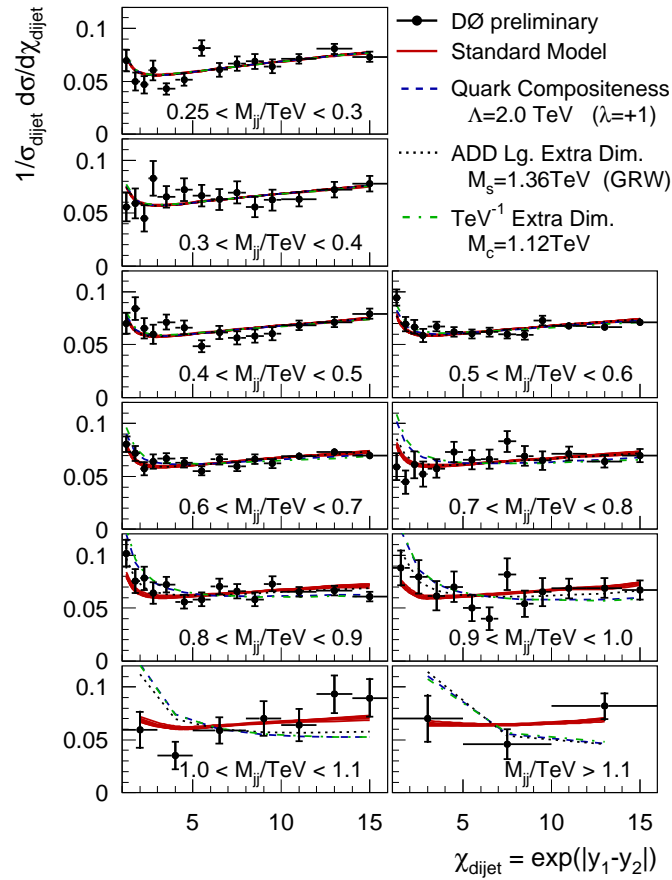}
\caption{$\chi_{dijet}$ angular distributions for data, QCD, and new physics models (D0).} \label{d0_dijet__3}
\end{figure*}

Jets are reconstructed by RunII midpoint cone 0.7 algorithm.
Data are corrected for insturmental effects such as detector
response, resolution, out-of-cone showering, additional energy,
vertex mis-identification, and jet reconstruction inefficiencies.
Dijet angular distributions at particle level are studied.
Dijet variable $\chi_{dijet} = exp(|y_1 - y_2|)$, where $y_1$, $y_2$
are rapidities of two leading jets, is directly sensitive 
to the dynamics of the underlying reaction. It is expected to have
different shape between QCD and new physics.
Particle level data distributions overlaid with QCD and several new physics
models are shown in Figure~\ref{d0_dijet__3}. Data do not show
significant deviations from QCD. Limits on three new physics models are
set using Bayesean apporach. Quark compositeness, ADD, and TeV$^{-1}$-sized Large Extra Dimensions (LED)
models are considered. The parameters of these models are energy scale $\Lambda$,
fundamental Planck Scale $M_S$, and  compactification scale $M_C$ respectively.
The observed limits on $\Lambda$, $M_S$, and $M_C$ are 2.58, 1.56, and 1.42 TeV. 

\section{$t\bar{t}$ SEARCHES}

\subsection{D0 $t\bar{t}$ Analysis}
An isolated electron with $p_T$ $>$ 20 GeV and $|\eta|$ $<$ 1.1 
or an isolated muon with $p_T$ $>$ 20 GeV and $|\eta|$ $<$ 2.0
is required. Missing transverse energy is required to be above
20 GeV (25 GeV) for electron (muon) channel. Three or more leading
jets, with one or more being b-tagged, are selected. $t\bar{t}$
invariant mass is reconstructed directly, without top quark mass constraint fit.
Main backgrounds are $t\bar{t}$, Z+jets, single top, dibosons, W+jets, and multijet.
Signal is modelled with the the high mass $Z^0$, whose width is equal to 1.2$\%$
of its mass. No resonant structure is observed. Limits are set
using Bayeasean approach. The results are shown in top left plot of
Figure~\ref{ttbar}. Leptophobic Z' is excluded up to 760 GeV.

\subsection{CDF $t\bar{t}$ Analysis with Template Method}
Event selection requires a central lepton with $p_T$ $>$ 20 GeV,
missing transverse energy above 20 GeV, and four jets with $|\eta|$ $<$ 2.0,
of which three must have $p_T$ $>$ 15 GeV, a fourth must have $p_T$ $>$ 8 GeV,
and at least one contains a secondary vertes b-tag. $t\bar{t}$ invariant
mass is reconstructed using the mass fit algorithm with top and W mass
constraints. Same signal model as described in previous section is used.
No evidence for resonances is observed. Limits are set using Bayeasean approach.
The results are shown in the top right plot of Figure~\ref{ttbar}.
For example, leptophobic Z' is ruled out below 720 GeV.

\subsection{CDF $t\bar{t}$ Analysis with Matrix Method}
Very similar event selection as in the previous analysis is used.
Matrix-element technique is used to reconstruct $t\bar{t}$ invariant
mass in each event. Same signal model as in previously described
$t\bar{t}$ analyses is used. In the absence of resonances, limits are set
using Bayeasean approach. The results are shown in the bottom left plot
of Figure~\ref{ttbar}. For example, leptophobic Z' is ruled out below 725 GeV.

\subsection{CDF $t\bar{t}$ Analysis with Dynamic Likelihood Method}
\begin{figure*}[t]
\centering
\includegraphics[width=85mm]{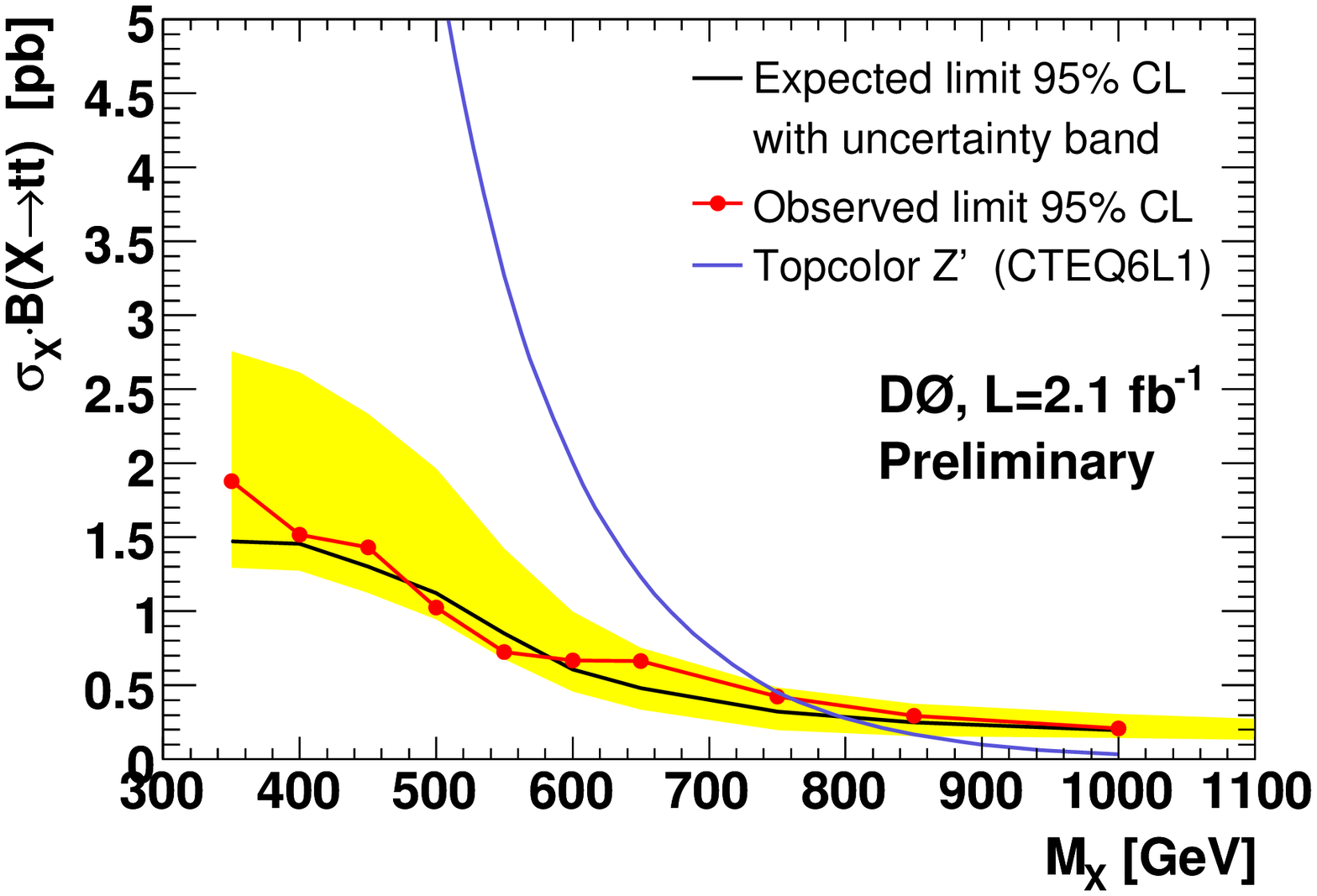}
\includegraphics[width=85mm]{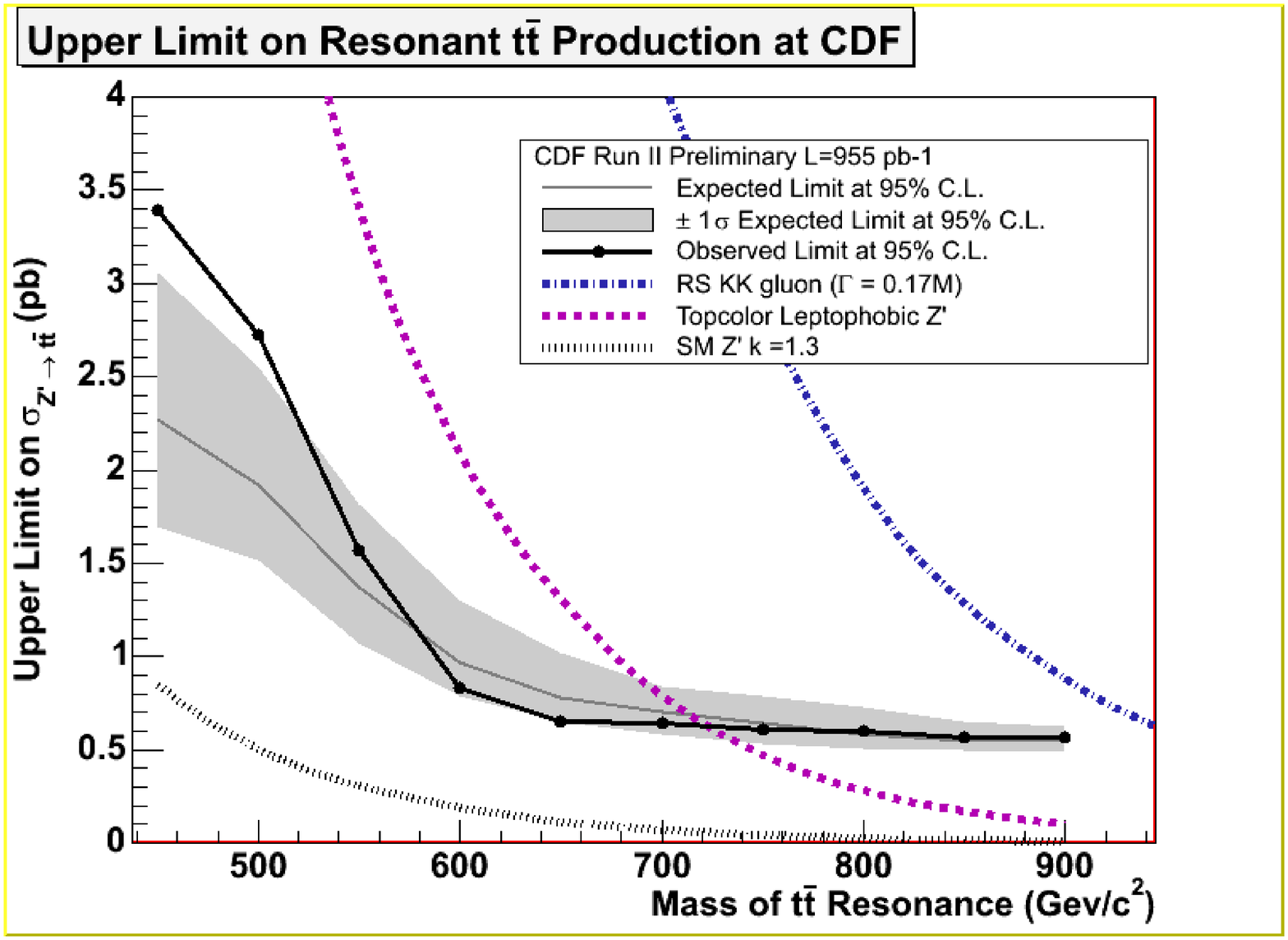}
\includegraphics[width=85mm]{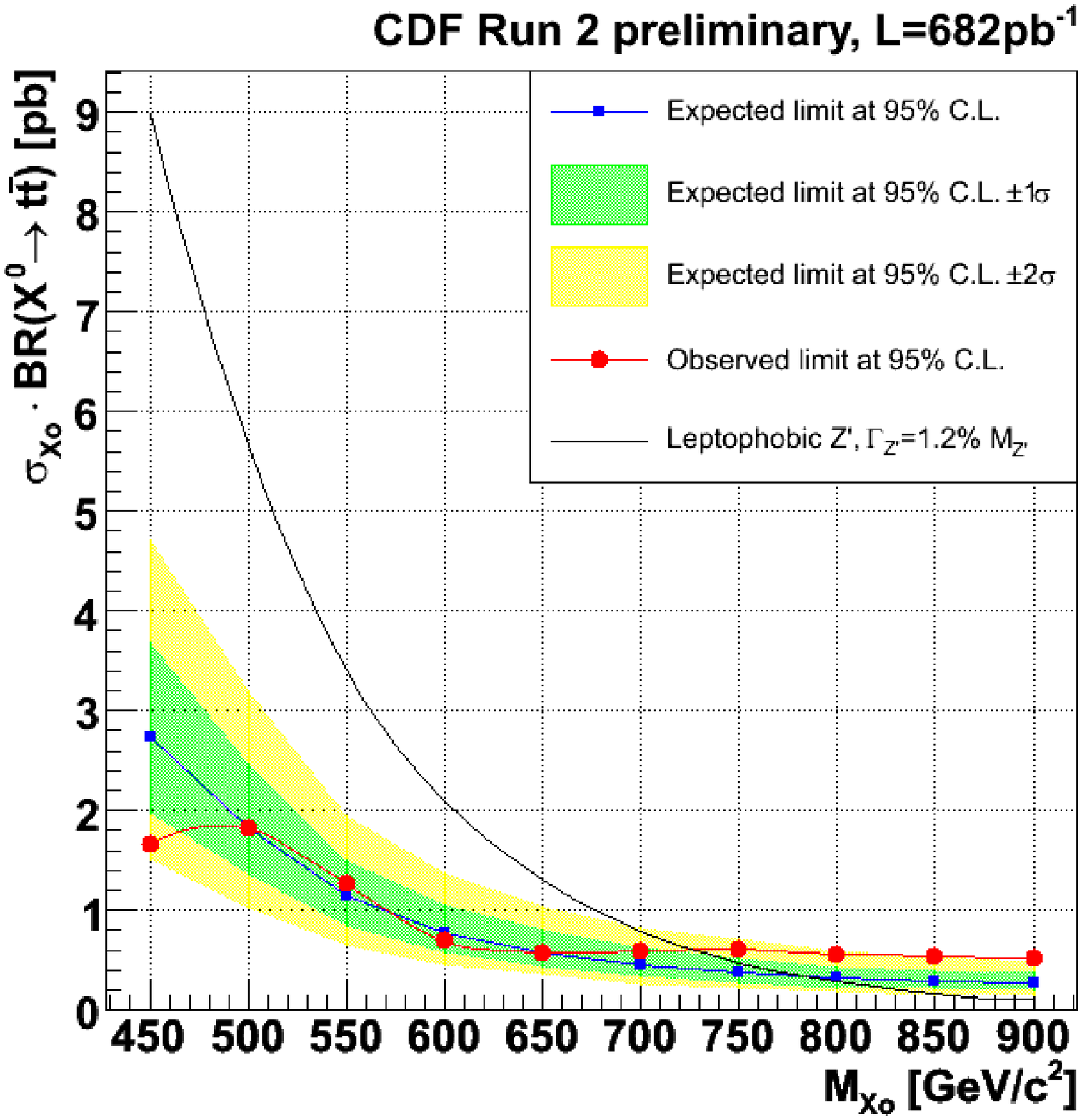}
\includegraphics[width=85mm]{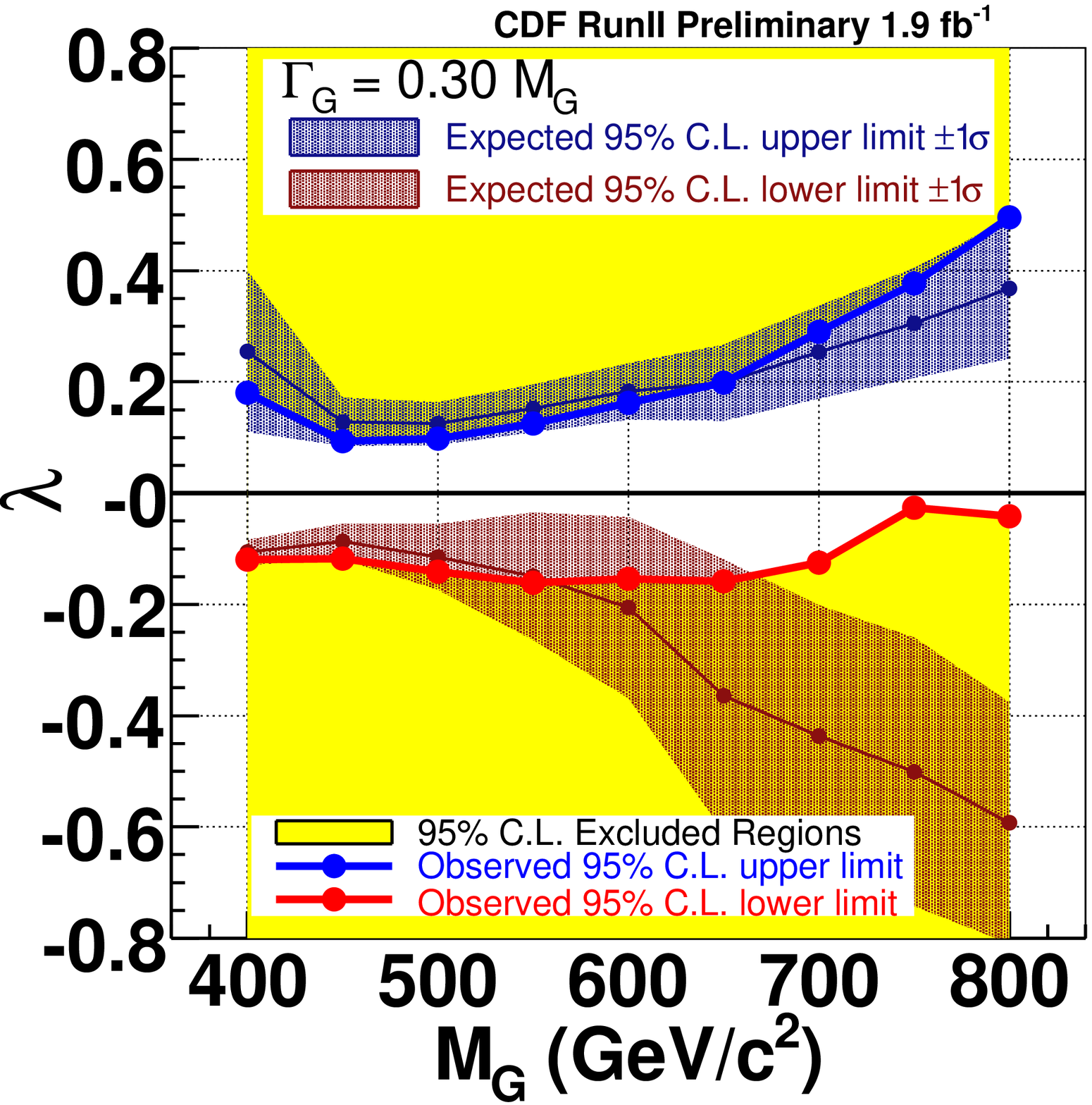}
\caption{$Results of t\bar{t}$ searches. Top left: D0 analysis. Top right:
CDF analysis with template method. Bottom left: CDF analysis with matrix method.
Bottom right: CDF analysis with Dynamical Likelihood method.
} \label{ttbar}
\end{figure*}

This is a search for the new color-octet particle called massive gluon.
The parameters are mass, width, coupling strengh. Mass range between
400 and 800 GeV is explored. Several width scenarios between 5$\%$
and 50$\%$ of the mass are considered. Exactly four jets with $p_T$ $>$ 
20 GeV and $\eta$ $<$ 2.0, reconstructed with the 0.4 cone algorithm 
are required. Muon or electron with $p_T$ $>$ 
18 GeV and missing transverse energy above 20 GeV are required.
The data are fit for the coupling strength and found to be consistent with the Standard Model.
Limits on the coupling strength are set. Exclusion contors for the case
when widht is equal to 30$\%$ of the mass are shownt on bottom right plot
of Figure~\ref{ttbar}. 

%

\begin{acknowledgments}
%
We thank the staffs at Fermilab and collaborating institutions, 
and acknowledge support from the 
DOE and NSF (USA);
CEA and CNRS/IN2P3 (France);
FASI, Rosatom and RFBR (Russia);
CNPq, FAPERJ, FAPESP and FUNDUNESP (Brazil);
DAE and DST (India);
Colciencias (Colombia);
CONACyT (Mexico);
KRF and KOSEF (Korea);
CONICET and UBACyT (Argentina);
FOM (The Netherlands);
STFC (United Kingdom);
MSMT and GACR (Czech Republic);
CRC Program, CFI, NSERC and WestGrid Project (Canada);
BMBF and DFG (Germany);
SFI (Ireland);
The Swedish Research Council (Sweden);
CAS and CNSF (China);
the Italian Istituto Nazionale di Fisica Nucleare;
the Ministry of Education, Culture, Sports, Science and Technology of Japan;
the Swiss National Science Foundation;
the AP Sloan Foundation;
the Comision Interministerial de Ciencia y Technologia (Spain);
the European Community's Human Potential Programme;
the Slovak R\&D Agency;
the Academy of Finland;
and the
Alexander von Humboldt Foundation (Germany).
\end{acknowledgments}


\begin{thebibliography}{9}   

\bibitem{cdf-prd} 
CDF Collaboration, T. Aaltonen {\it et al.},
Phys. Rev. D {\bf 77}, 051102(R) (2008).

\bibitem{cdf-prl} 
CDF Collaboration, T. Aaltonen {\it et al.},
Phys. Rev. Lett. {\bf 100}, 231801 (2008). 

\end{thebibliography}
\end{document}